# Deepfake Labels Restore Reality, Especially for Those Who Dislike the Speaker

Nathan L. Tenhundfeld[1], Ryan Weber[1], William I. MacKenzie[1], Hannah M. Barr[2], Candice Lanius[1]

[1]University of Alabama in Huntsville

[2]Communication and Culture, LLC

## Abstract

Deepfake videos create dangerous possibilities for public misinformation. In this experiment (N=204), we investigated whether labeling videos as containing actual or deepfake statements from US President Biden helps participants later differentiate between true and fake information. People accurately recalled 93.8% of deepfake videos and 84.2% of actual videos, suggesting that labeling videos can help combat misinformation. Individuals who identify as Republican and had lower favorability ratings of Biden performed better in distinguishing between actual and deepfake videos, a result explained by the elaboration likelihood model (ELM), which predicts that people who distrust a message source will more critically evaluate the message.

## Public Significance Statement

Deepfake videos depict events that never happened, such as politicians making statements they never actually made, and pose a significant threat to misinform the public. As deepfake technology continues to improve, videos may become even more convincing and pose a greater threat. Therefore, it is imperative to explore methods of mitigating their ability to spread misinformation. This research tested whether labels help people remember whether information in a video is real or fake. People frequently remembered whether information was actual or deepfake based on labels about information accuracy, suggesting that labeling actual and deepfake information can combat misinformation.

## Introduction

Deepfake videos are created using artificial intelligence to realistically present events that never happened, and pose serious problems for social trust, democratic deliberation, personal relationships, and national security (Pawelec, 2022; Mustak et. al., 2023). The heart of this threat comes from an audience's inability to determine and remember whether videos depict real or fictitious events. Deepfake videos can show politicians making statements they never made or invent and alter news stories, creating the possibility of an "epistemic threat" (Fallis, 2021) where the public has less confidence in the information it receives.

Existing research into deepfakes shows that people often struggle to detect them (Khodabakhsh, Ramachandra, & Busch, 2019; Hameleers et. al., 2023). Almost half of participants considered deepfake videos credible, especially when the videos aligned with existing political views (Barari et. al., 2021). Some research finds audience ability to detect deepfakes at between 59% and 69% depending on video quality (Rosseler et. al., 2019) while a more recent study has shown deepfake recognition as low as 40% (Appel & Prietzel, 2022). As this technology continues to progress, the ability to discern between authentic and deepfake videos should only become more challenging. Even when given warnings about the existence of deepfakes and financial incentive to detect them, people often struggle to identify deepfakes, despite an overconfidence in their abilities to do so (Kobis et. al., 2021). People tend to assume others will be more influenced by deepfakes than themselves, but this belief does not improve detection ability (Ahmed, 2021).

Regardless of whether deepfakes have been identified by an audience as fake, they may still lead to adverse outcomes. Audiences can be manipulated by videos even when they know that they are fake (Hughes et. al., 2021). For example, repeated exposure tends to make deepfakes more credible (Deira, 2023). Negative deepfake videos of politicians can lower audience perceptions of them (Dobber et. al., 2021), and even uncredible deepfakes can negatively influence perceptions of politicians (Hameleers & Marquart, 2023).

Despite the concerns about deepfakes and audience inability to detect them, researchers have not conclusively established that deepfakes misinform more effectively than other media (Murphy & Flynn, 2022). Participants are slightly more likely to believe deepfake videos over text describing similar misinformation (Wittenberg et. al., 2021), but in many cases deepfake videos may not spread misinformation more effectively than text (Hamleers et. al., 2022; Murphy et. al., 2023), text and pictures (Murphy & Flynn, 2022), or visuals without audio (Xu et. al., 2020). People may also mistrust decontextualized accurate information as much as disinformation (Hameleers et. al., 2022). However, politicians have more success suggesting text-based news is false than when claiming video-based news is false (Schiff, Schiff, & Bueno, 2023), suggesting that participants sometimes regard videos as more credible than text.

Many venues have adopted an approach to counter deepfakes in which mis- and disinformation are labeled to inform the consumer, at the time of consumption, that the information is fake. While venues may hope that labels help people remember which information is true, people may become confused later when trying to remember which sources were accurate, or they may simply ignore warning labels. This sort of “source confusion” or “source misattribution” (Zaragoza & Lane, 1994) wherein an individual remembers seeing a deepfake video but thinks that it was an actual video, or vice versa, poses a potential risk for the propagation of disinformation.

Research demonstrates that some informational mitigation efforts can help audiences identify deepfakes and protect against deception. Corrective labels may keep audiences from sharing deepfakes (Ahmed, 2021). Messages and countermeasures against deepfakes help, but only to a point (Barari et. al., 2021). Although warnings about deepfakes doubled people’s ability to successfully identify them, almost 80% of people still cannot accurately identify deepfakes (Lewis et. al., 2022). Teaching participants strategies for detecting deepfakes does not improve their performance (Somoray & Miller, 2023), though giving people media literacy education can reduce the effectiveness of deepfakes (Hwang, Ryu, & Jeong, 2021). However, the very existence of deepfakes reduces trust in news overall (Vaccari & Chadwick, 2020). Warnings about deepfakes cause participants to assume that all videos are fake (Ternovski, 2022), and even groups that accurately identified deepfakes on average still tended to assume that real videos were fake (Ng, 2022). Therefore, even when deepfakes do not directly misinform, they make audiences more skeptical generally.

The problem of misinformation may be worse by people’s tendency to more eagerly embrace misinformation that aligns with their ideology (Traberg & Linden, 2021; Shin & Lee, 2021). Research from ethnographic interviews suggests that “people are susceptible to fake videos that affirm their political or communal biases” (Shahid, et. al., 2022, p. 4). Participants with out-partisan identification (in this case, identification against liberal US political figures) found deepfakes slightly more credible, though Republicans often outperformed Democrats in identifying deepfake videos (Barari et. al., 2021). US conservatives may be more susceptible to misinformation than US liberals (Ecker & Ang, 2019; Garrett & Bond, 2021; Baptista & Gradim, 2022; Gupta et. al., 2023). A small subset of carefully micro-targeted voters can have their attitudes influenced by political deepfakes about their own candidate, though many people will engage in motivated reasoning to innoculate themselves against the effects of deepfakes (Dobber et. al., 2021). General right-wing political orientations do not increase participants' credibility ratings of deepfakes espousing extreme right-wing statements, though participants with strong anti-immigration stances were more likely to perceive disinformation related to anti-immigration as credible (Hameleers, 2022). Other research indicates a small correlation between social conservatism and a tendency to perceive deepfake videos as authentic (Sütterlin et. al., 2023). Political interest does not always correlate with ability to identify deepfakes (Appel & Prietzel, 2022), but participants with high political interest are more likely to unwittingly share deepfakes (Ahmed, 2021). General political knowledge can improve participants' resistance to misinformation and distortion (Calvillo & Rutchick, 2014).

However, because deepfakes are new technology, we still do not fully understand how people perceive and respond to them. Scholars have called for more research into how audiences react to deepfakes (Mustak et. al., 2023) as well as strategies to mitigate the effect of politically damaging deepfakes (Godulla et. al., 2021). This experiment investigates whether labeling improves an individual's ability to remember if a video contains actual or fake information. The researchers presented participants with either video of current president Joe Biden that contained either actual statements made by Biden or statements created with deepfake technology. Each video was accompanied by a corresponding label that indicated whether the video contained a real or deepfake statement. Participants were then asked to recall whether statements from the videos were actually made by Biden or created as deepfakes. Overall, participants showed excellent recall for labeled content, especially in their ability to identify deepfake statements. Participants who had negative favorability ratings for President Biden performed especially well in identifying actual and deepfake videos, a finding that aligns with predictions based in the elaboration likelihood model (ELM) that individuals who distrust a message source will engage more carefully with message content.

**Elaboration Likelihood Model**

The elaboration likelihood model (ELM) is a useful framework for understanding how an audience might interact with deepfake videos. The ELM is a dual process model of persuasion where persuasiveness is dependent on the degree to which message receivers will elaborate on the information they receive (Petty & Cacioppo, 1986). In the context of the ELM, elaboration is defined as individuals engaging in careful thought about information received in a persuasive message (Fleming & Petty, 1999). Some messaging will trigger higher levels of elaboration, which results in systematic thinking. When elaboration is higher, individuals engage more thoroughly with the information they are given and are more likely to carefully examine the message’s content.

Persuasive messages that are associated with lower levels of elaboration are more likely associated with heuristic decision making. Typically, lower levels of elaboration occur when messages come from individuals who the message receiver views as credible or likable (Petty & Cacioppo,1986; Sinclair et al., 2010). When message receivers like the communicator, they are less likely to thoroughly engage with the content of the message (low elaboration) and subsequently rely on heuristics in their decision making. Alternatively, viewing a message from a source that is disliked by the receiver should result in increased levels of elaboration which leads to the receiver focusing more on the content of the message. In the ELM framework persuasion can occur when thought is high or low but the processes are very different in each condition (Petty & Wegener, 1999). In the current study, ELM allows us to understand how message receivers' attitudes toward the message source trigger different levels of cognition among viewers.

While the ELM has historically been used to explain the persuasiveness of messages, we find it to be a useful framework in explaining the insidious nature of malicious deepfake videos. ELM has been used as a framework for examining political communication (e.g., Chebat et al., 1990). The ELM suggests when individuals engage in high elaboration they will carefully evaluate the information they are given (Petty et al., 2009).

There are multiple aspects of a message that might influence an individual's level of motivation to think about a message (Petty et al., 2009). The source of a message is a common example of an element of a message that impacts elaboration heuristics. (Chaiken, 1987). For example, information from a trusted source is often accepted by message receivers without the need for elaboration. The opposite has also been demonstrated, where a questionable or untrustworthy source has been shown to increase elaboration among message receivers even when the source's expertise was manipulated to be high (Priester & Petty, 1995). Across studies that manipulated source trustworthiness, sources who were viewed as less trustworthy triggered greater elaboration than those who were viewed as more trustworthy (Petty et al., 2009). This effect is so strong that it can trigger elaboration from individuals who typically would not engage in elaboration (Cacioppo & Petty, 1982). The ELM provides an interesting theoretical lens for examining deepfake videos which by design manipulate the source of messages received by the viewer.

**Purpose and Research Questions**

This study examines how well participants can recall whether political statements came from authentic or deepfake videos. Participants watched a series of 52 videos of current US President Joe Biden that contained both authentic statements from Biden and deepfakes created by the researchers. Each video was accompanied by text underneath it that labeled the video as either actual or a deepfake. After watching the videos, participants reviewed a list of statements that appeared in the videos and identified whether the statements appeared in a deepfake or authentic video. Participants also evaluated how believable it was that President Biden would say each of the deepfake statements.

This study expands on previous research regarding deepfake detection by examining whether participants can correctly remember whether statements came from authentic or deepfake sources. Although several studies have examined whether participants can correctly identify deepfake videos (Khodabakhsh, Ramachandra, &

Busch, 2019; Hameleers et. al., 2023) and the influence of warning labels on audience intentions to share deepfake videos (Ahmed, 2021), very little research has examined how well people recall labeled information and misinformation. The researchers created original deepfake videos about a variety of topics, responding to calls for research that employs never-before-seen deepfakes about multiple subjects (Appel & Prietzel, 2022), as many experiments use previously created deepfake videos that participants may have already seen.

This study answers the following research questions:

1. Can participants accurately recall whether statements came from videos labeled as actual or deepfake?
2. Does participant ability to recall whether statements come from actual or deepfake videos vary as a function of political difference (political party affiliation, political identification, and political ideology) and attitudes towards President Biden?
3. Does participant ability to recall whether statements came from actual or deepfake videos vary as a function of news and social media consumption habits?

## Methods

### Participants

Participants were recruited through MTurk and available Psychology courses at a medium sized university in the southeastern United States. Data from 204 participants was collected. Participants had a mean age of 21.6 years (*SD* = 6.08). Approximately 52% of participants were female (44.6% male, 3.4% other/prefer not to say).

Participants were 70.6% white, 9.4% black, 3.9% Asian, 1.5% hispanic/latino, 1.5% native american, and the rest were mixed/multi-racial. Of our participants, 36.3% identified as Democrats, 36.3 identified as Independents, and 27.5% identified as Republicans.

**Stimuli & Materials**

*Videos.* Deepfakes were generated by using the First Order Motion Model for Image Animation (FOMM-IA) technique developed by Fross et al., (2019), coupled with the “Celebrity Voice Changer Parody” (CVCP) application through the Apple Store. The lead author recorded a clip of himself saying the deepfake statements with the CVCP application, which then altered his voice to sound like President Joe Biden. The videos were then downloaded (with the audio). The FOMM-IA approach animates a static image using input from an existing video. As shown in Figure 1, the resultant animation mirrors the head and mouth movements detected in the existing video. We collected static images of President Biden and created an animation of every static image for every existing video. We kept the most realistic animation for every existing video and synced it with the CVCP audio.

**Figure 1**

*Process for deepfake video creation.*

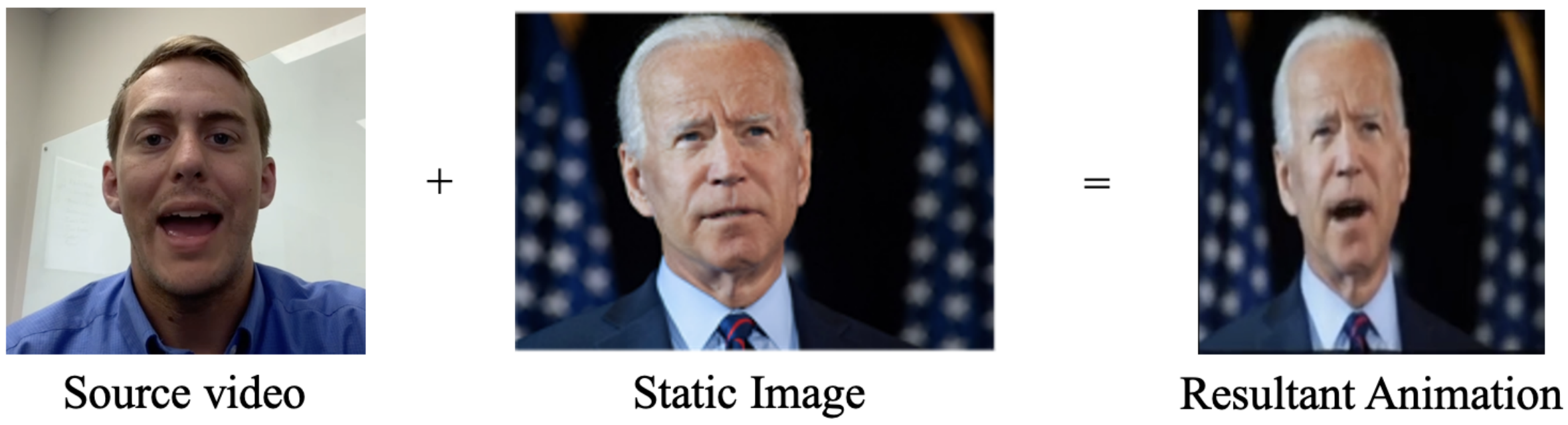

Actual videos were selected which depicted President Biden talking about topics which mirrored topics covered in the deepfake videos. A list of statements used for the deepfakes and their real-video counterparts can be found in Table X. The deepfake statements were designed to range from the plausible (i.e. “Americans should be wearing masks until at least 2023”) to outlandish/extreme (“I would be in favor of changing the name of the White House to something more politically correct”) (See Appendix A for all video statements). All videos were cropped to 256x256 pixels to equate resolution.

*Surveys*

We employed several surveys to assess the role of individual differences in the participant’s performance on label recall. The list of surveys can be found in Table 1. Before the presentation of the videos, participants indicated how much they liked Joe Biden on a scale of “dislike a great deal" to “like a great deal”. Participants also completed the source credibility scale (McCroskey & Teven, 1999) for their impressions of Joe Biden by selecting points between dichotomies. Such dichotomies used were “honest/dishonest", “phony/genuine", and “insensitive/sensitive", among others.

After the presentation of the videos, participants then completed a series of measurements of individual differences. These measurements began with the PANAS (Watson et al., 1988), used to measure their current levels of positive and negative affect. This scale required participants to rate their affect based on certain words such as “distressed", “excited", “proud", “nervous", and more on a scale of “very slightly or not at all” to “extremely" based on how they felt in the moment. This scale was followed by a media consumption scale (Vaccari & Chadwick, 2020) used to ascertain participants’ media habits. In this scale, participants selected on a scale of “not at all" to “ a great deal" (with an “I don’t know” option) how much they trusted the news they see on social media. They also indicated how many hours a day (0-12) they spend on social media and how many hours they spend watching or reading news. Additionally, participants were asked to indicate where they get their news. Participants also provided their political party affiliation (Greene, 2004) with American parties by selecting either Republican, Democrat, or Independent. This question was followed by participants responding to a series of items regarding their level of political identification with the party on a scale of “strongly disagree" to “strongly agree"; items included statements such as “I don’t act like the typical person of this group", “the group’s successes are my successes", and “I’m very interested in what others think of this group”, among others. The survey also collected personality measures based on the Need for Identification Scale (Mayhew et. al., 2010), the Personality Item Pool (IPIP) A1: Trust scale (Goldberg et. al., 2006), and the conspiracy mentality scale (Bruder et al., 2013) all of which were measured on a scale of “strongly disagree” to “strongly agree."

**Table 1**

*Surveys used in present study*

| **Survey** | | **Citation** | **Measure** | |
|---|---|---|---|---|
| Views on President Biden | 7-point Likert Scale | | Liking of President Biden | Pre- & Post-Videos |
| Source Credibility/Goodwill | 7-point Bipolar Scale | (McCroskey & Teven, 1999) | Competence, Goodwill, & Trustworthiness | Pre-Videos |
| Scrutiny | 4-point Likert Scale | (Dobber et al., 2021; Wheeler et al., 2005) | Engagement with videos | Post-Videos |
| PANAS | 5-point Likert Scale | (Watson et al., 1988) | Positive and Negative Affect | Post-Videos |
| Trust in Social Media | 4-point Likert Scale | (Vaccari & Chadwick, 2020) | Trust in news and information about politics and public affairs seen on Social Media | Post-Videos |
| Identification | 5-point Likert Scale | (Greene, 2004; Mael & Tetrick, 1992) | Level of Identification with Selected Political Party | Post-Videos |
| Need for Identification | 5-point Likert Scale | (Mayhew et al., 2010) | Propensity to identify with social groups | Post-Videos |
| Trust – IPIP | 5-point Likert Scale | (Goldberg et al., 2006) | Propensity to trust others | Post-Videos |
| Extraversion – IPIP | 5-point Likert Scale | (Goldberg et al., 2006) | Trait extraversion | Post-Videos |
| Conscientiousness – IPIP | 5-point Likert Scale | (Goldberg et al., 2006) | Trait conscientiousness | Post-Videos |
| Conspiracy Mentality Questionnaire | 5-point Likert Scale | (Bruder et al., 2013) | | |
| Better Than Average Effects | | (Moore, 2007) | Belief in one's own ability | |

## Procedure

This study was administered both online and in person in a research lab setting from August 2021 through April 2023 using Qualtrics. While these two different delivery methods were used, nothing else regarding the delivery of the study varied. The study, which was approved by the institutional research bureau at [Institution name withheld] began by collecting informed consent from participants. Following this, an audio/visual check was conducted to determine whether participants could see and hear the videos being presented. Participants also completed attention checks to ensure they paid attention to the content of the audio/visual check video. If participants failed either the audio/visual check or the attention check, the study ended.

For participants who passed these checks, the study continued by having participants respond to a set of items regarding their current views towards Joe Biden. Participants indicated how much they liked Joe Biden on a scale of “dislike a great deal" to “like a great deal”. Participants also described their impressions of Joe Biden by selecting points between dichotomies. Such dichotomies used were “honest/dishonest", “phony/genuine", and “insensitive/sensitive", among others. Following these measures, participants were instructed to watch a series of 51 clips of Joe Biden, some of which were real videos and some of which were deepfakes. Each video had a label presented below it reading either “actual video” or “Deepfake" according to whether the video was authentic or researcher-generated. These videos lasted between 2 and 14 seconds each and covered various topics, such as addiction, Biden stepping down as president, border policy, tax policy, and more (See Table 2 for video topics). Once participants finished watching all of the video clips, they responded to a set of items to indicate how attentively they watched the videos, the extent that they thought deeply about the content of the videos, and how much effort they put into understanding the content of the videos. Responses to the could range from “not at all" to “a great deal", with an additional option to reply “I don’t know”. At this point, participants also labeled a series of statements based on whether or not they remembered hearing them in the video.

Participants then completed surveys to measure personality and individual differences. Following these individual difference measures, the survey collected several types of demographic information. This information included age, where one lives (urban, suburban, or rural), education, income, race, and gender. Participants were also asked to indicate their political ideology by placing themselves on a seven-point scale from “Extremely Liberal” to “Extremely Conservative” (ANES, 2024). Participants were also asked again to indicate their current views towards Joe Biden on a scale of “dislike a great deal" to “like a great deal".

At this point in the study, participants were asked to consider the videos they had viewed earlier. The statements from each video were listed and participants were tasked with recalling whether the videos were labeled as actual or as deepfake videos. Participants were then tasked with rating how well they think they performed in recalling labels and how they believed their performance at recalling the labels would compare to other Americans, using items adapted from Lyons et al., (2021). Following this, participants were presented with all of the statements used in the deepfake videos and were asked to rate their believability on a scale of “extremely unbelievable” to “extremely believable". Participants were then asked to take a working memory digit span task (Ryan & Lopez, 2001) to conclude the study. The task began with 3 digits and increased to 9 digits maximum, chosen due to Miller's (1955) idea that working memory can hold 7 plus or minus 2 items. Once participants missed two sets of numbers in a row, the working memory task ended. At the end of this task, participants were debriefed, thanked for their time, and released from the study.

**Analyses**

In addition to other frequentist and bayesian analyses reported below, we used signal detection measures to assess performance. Within signal detection theory, sensitivity (d’) can be calculated to ascertain how good people are at discriminating between the distributions of stimuli (in this case the deepfake videos and the actual videos). A greater value for d’ indicates greater discriminability. Sensitivity scores (d’) were calculated for participants. However, d’ cannot be calculated for those with a perfect hit or false alarm rate, and as such we adjusted hit and false alarm rates for those with perfect rates, in line with Macmillan & Kaplan (1985). For participants with a hit rate of 1.0, we subtracted 0.5 from their hits (27), and divided it by the number of possible hits. This meant that their hit rates were given a value of 0.981. Similarly, participants with a false alarm rate of 0 were given a FA rate of 0.021 as this corresponded to .5 divided by the number of possible correct rejections. Next, we calculated response bias with B", such that a score of 0 indicates no response bias. More negative numbers indicate a greater response bias towards recalling a video as being a deepfake, whereas more positive number indicate a greater response bias towards recalling a video as being an actual video. In other words, B" is a measure of the strength and direction of errors that participants make.

## Results

Recall performance was calculated as a proportion of correctly recalled labels for each statement associated with the videos they watched. A Repeated Measures ANOVA with video type (deepfake versus actual) as the within subjects variable and self-identified political ideology as the between subjects factor revealed a significant main effect between proportion of deepfake videos correctly recalled as deepfake ($M$ = 0.938, $SD$ = 0.113) and actual videos correctly recalled as actual videos ($M$ = 0.842, $SD$ = 0.141), $F(1, 197) = 52.516$, $p < .001$, $\eta^2 = 0.063$.

There was no main effect between self-identified political ideology for recall performance, $F(6, 197) = 1.359$, $p = 0.233$, $\eta^2 = 0.028$. There was also no interaction between recall performance by video type and self-identified political ideology, $F(6, 197) = 0.873$, $p = 0.516$, $\eta^2 = 0.006$. Figure 2 shows the proportion of videos correctly identified, by video type, and political ideology. A Bayesian Repeated Measures ANOVA indicated that the most likely model was the one with only a main effect for video type, $BF_{10} > 1,000$. However, this model was only 1.11 times likelier than a model that included a main effect for political ideology as well ($BF_{10} > 1,000$). A repeated measures ANOVA using political party affiliation as the between subjects variable indicated no significant difference between political party affiliation on recall performance, $F(2, 201) = 1.574$, $p = 0.210$, $\eta^2 = 0.010$. Additionally there was no interaction between video type and political party affiliation on recall performance, $F(2, 201) = 1.518$, $p = 0.222$, $\eta^2 = 0.003$. Once again, this finding was supported by a Bayesian Repeated Measures ANOVA which indicated that the most likely model was the one with only a main effect for video type $BF_{10} > 1,000$.

**Figure 2**

*Proportion of videos correctly identified, by video type, and political ideology.*

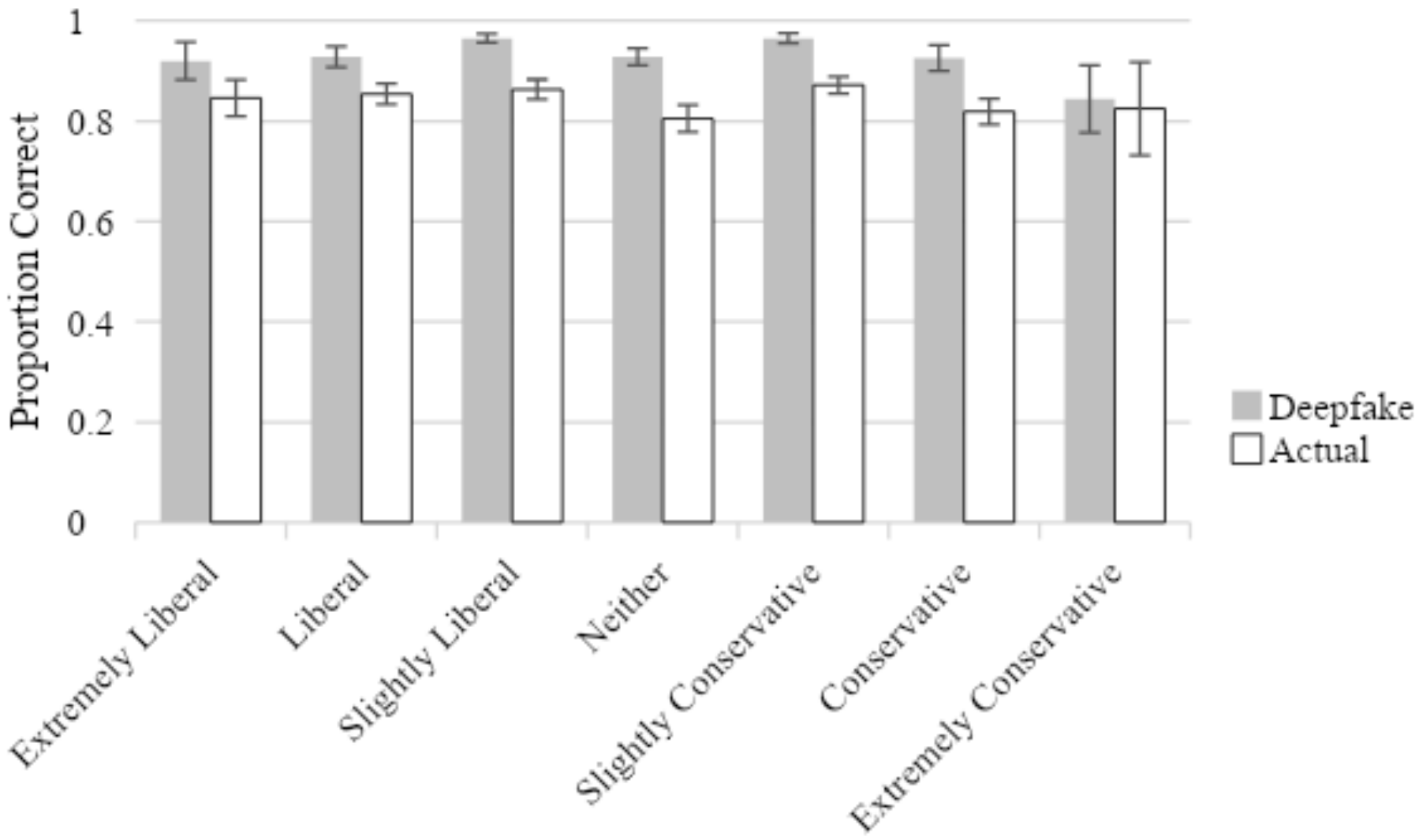

*Note.* Error bars represent ±1 SEM.

**Attitudes towards Biden**

A separate Repeated Measures ANOVA with video type as the within-subjects variable and self-reported attitudes towards President Biden as the between subjects variable again indicated a significant main effect for video type on recall performance $F(1, 197) = 50.926$, $p < .001$, $\eta^2 = 0.059$. There was no significant main effect for attitudes towards President Biden on recall performance, $F(6, 197) = 2.050$, $p = 0.061$, $\eta^2 = 0.041$. There was also no significant interaction between video type and attitudes on recall performance, $F(6, 197) = 1.906$, $p = .082$, $\eta^2 = 0.013$. Figure 3 shows the proportion of videos correctly identified by video type and participant views toward President Biden. However, a Bayesian Repeated Measures ANOVA indicated the most likely model is the one with a main effect for both video type and views of President Biden ($BF_{10} > 1{,}000$), which was 2.990 times likelier than a model with just a main effect for video type.

**Figure 3**

*Proportion of videos correctly identified, by video type, and views on President Biden*

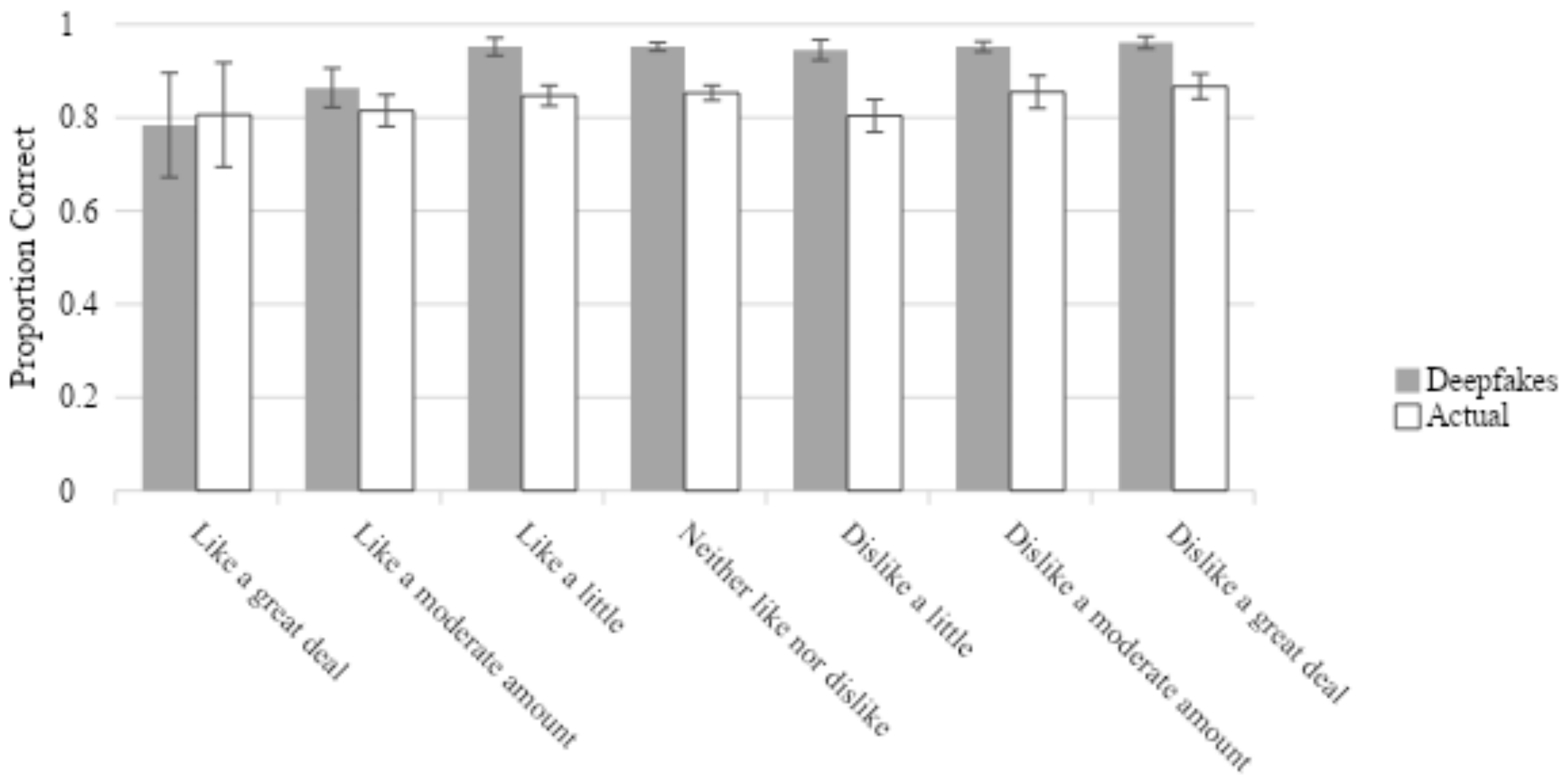

*Note.* Error bars represent ±1 SEM.

Recall performance on both the deepfake and actual videos were regressed upon the change in views on President Biden from before to after the videos were shown. Figure 4 illustrates participant views towards President Biden before and after viewing videos. The regression was not significant, $F(2, 201) = 0.109$, $p = 0.896$, and neither actual video recall performance ($p = 0.647$, $BF_{10} = 0.162$), nor deepfake video recall performance ($p = 0.761$, $BF_{10} = 0.153$) was a significant predictor of view change. There was no significant difference on self-reported attitudes on President Biden between ratings given before watching the videos ($M = 3.681$, $SD = 1.554$) and after watching the videos ($M = 3.765$, $SD = 1.574$), $t(203) = 1.792$, $p = 0.075$, $BF_{10} = 0.376$.

**Figure 4**

*Views on President Biden before and after exposure to deepfake and actual videos*

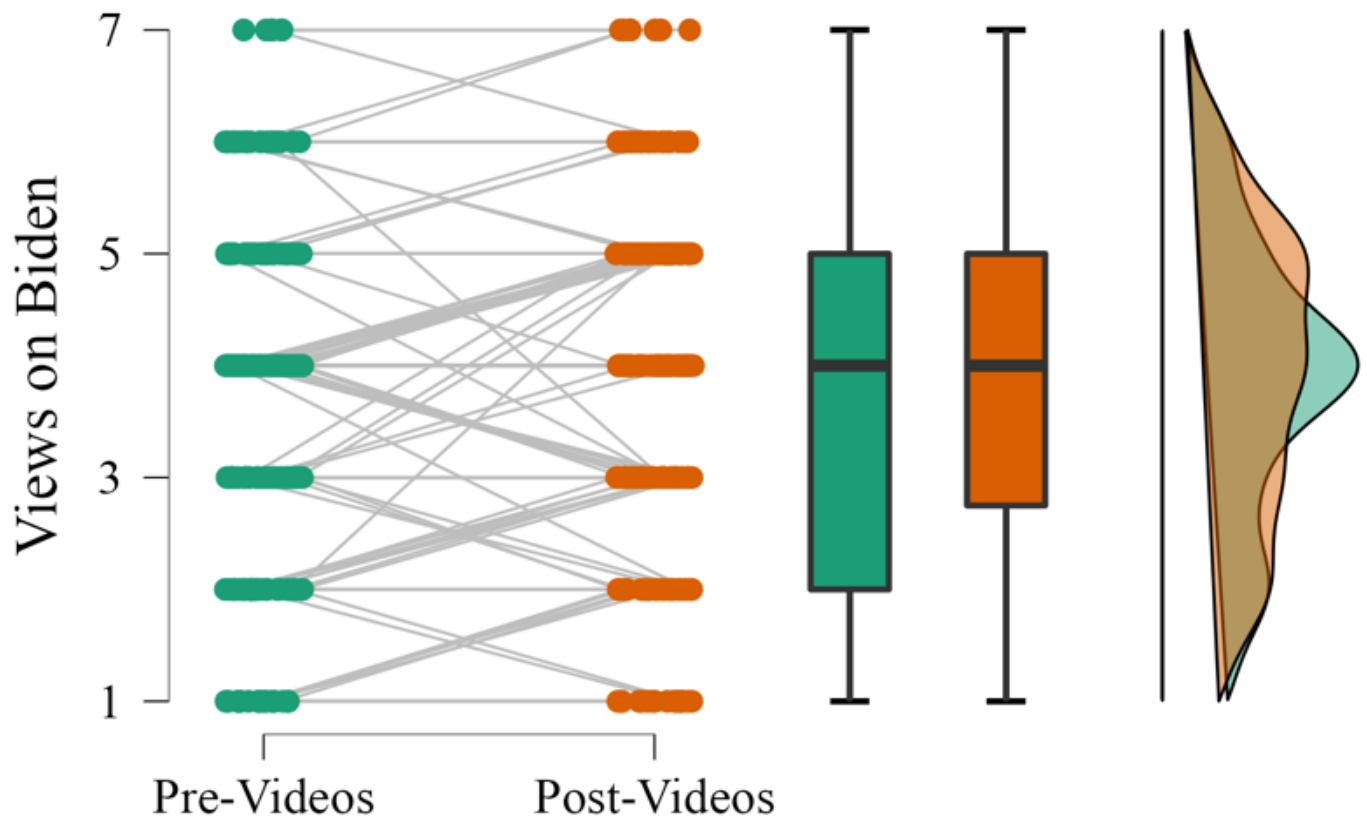

*Note*. Each line connects an individual's views on President Biden given pre- and post-videos. Boxplot and raincloud plot depict the distribution.

## WMC & Recall Performance

There was no significant correlation between working memory capacity and actual video recall performance, $r(202) = 0.106$, $p = 0.131$, $BF_{10} = 0.272$. There was no significant correlation between working memory capacity and deepfake video recall performance, $r(202) = 0.084$, $p = 0.233$, $BF_{10} = 0.177$. A surprising number of participants ($N = 149$) had a score of 0 for the measure of their WMC. When those participants were excluded, however, the results did not change.

**Prediction of Recall Performance**

Because of the plethora of individual differences predictors, we submitted the variables found in Table 3 to a stepwise linear regression to predict recall performance on the deepfake videos. Criteria for entry was set at $p < .05$, while removal was set at $p > .10$. The best fit model included three negative predictors (i.e. as these went up, performance went down): hours spent per day reading and watching the news, the believability of the deepfake videos, views on President Biden (pre-videos), and one positive predictor (i.e. as this went up, so too did performance): the number of sources someone accesses to get his/her news (see Table 4). This model was a significant predictor of recall performance, $F(4, 194) = 33.049$, $p < .001$, $R^2 = 0.405$.

**Table 3**

Individual Difference Predictors Tested

| **Personality** | **Behavioral** | **Political & Paradigm Specific** |
|---|---|---|
| • PANAS Positive Affect<br>• PANAS Negative Affect<br>• Trust in News and Information on Social Media<br>• Trust in others<br>• Need for Group Identification<br>• Extraversion<br>• Conscientiousness<br>• Conspiracy Mentality<br>• Working Memory Capacity | • Hours Reading/Watching News per Day<br>• Hours on Social Media per Day<br>• Number of Sources for Media Consumption | • Believability of Deepfakes<br>• Views on President Biden (Pre-videos)<br>• Political Ideology<br>• Source Credibility<br>• Party Belonging |

**Table 4**

Stepwise linear regression on deepfake recall

| Model | $R^2$ | | β | t | p |
|---|---|---|---|---|---|
| 1 | 0.000 | (Intercept) | | 118.727 | < .001 |
| 2 | 0.189 | (Intercept) | | 98.519 | < .001 |
| | | Hours Reading/Watching | -0.435 | -6.779 | < .001 |
| 3 | 0.326 | (Intercept) | | 56.241 | < .001 |
| | | Hours Reading/Watching | -0.388 | -6.570 | < .001 |
| | | Believability | -0.373 | -6.308 | < .001 |
| 4 | 0.392 | (Intercept) | | 45.123 | < .001 |
| | | Hours Reading/Watching | -0.338 | -5.888 | < .001 |
| | | Believability | -0.439 | -7.555 | < .001 |
| | | Views on Biden | -0.269 | -4.603 | < .001 |
| 5 | 0.405 | (Intercept) | | 43.208 | < .001 |
| | | Hours Reading/Watching | -0.387 | -6.279 | < .001 |
| | | Believability | -0.431 | -7.464 | < .001 |
| | | Views on Biden | -0.266 | -4.601 | < .001 |
| | | Number of Sources | 0.125 | -2.075 | .039 |

The same predictor variables were submitted to a stepwise linear regression upon actual video recall performance. In this case, the best fit model was the one which included only the hours spent reading and watching the news and believability of the deepfake videos $F(2, 198) = 20.185$, $p < .001$, $R^2 = 0.171$ (see Table 5). Both predictors showed a negative relationship such that higher believability ratings and more hours spent reading/watching the news resulted in lower recall performance. While participants did not provide believability scores for the actual videos, we reasoned that it was appropriate to keep believability scores for this model as the more believable the deepfake videos were, the harder it would be to parse the actual videos from the deepfakes.

**Table 5**

Stepwise linear regression on actual video recall

| Model | $R^2$ | | β | t | p |
|---|---|---|---|---|---|
| 1 | 0.000 | (Intercept) | | 83.613 | < .001 |
| 2 | 0.189 | (Intercept) | | 66.885 | < .001 |
| | | Hours Reading/Watching | -0.343 | -5.121 | < .001 |
| 3 | 0.326 | (Intercept) | | 35.491 | < .001 |
| | | Hours Reading/Watching | -0.314 | -4.784 | < .001 |
| | | Believability | -0.233 | -3.550 | < .001 |

**Sensitivity and Bias**

Participants' sensitivity (d') ($M$ = 2.841, $SD$ = 0.924) was significantly greater than 0, $t(203) = 43.894$, $p < .001$, $BF_{10} > 1000$. Indicating that they were significantly better than chance at differentiating between the deepfake and actual videos. There was no significant difference between different self-reported political ideology and sensitivity, $F(6, 197) = 1.310$, $p = .254$, $\eta^2 = 0.038$, $BF_{10} = 0.121$. Similarly, there was no significant difference between self-reported views on President Biden and sensitivity, $F(6, 197) = 1.533$, $p = .169$, $\eta^2 = 0.045$, $BF_{10} = 0.140$. There were no significant differences between political party affiliation on sensitivity, $F(2, 201) = 1.510$, $p = .223$, $\eta^2 = 0.015$, $BF_{10} = 0.195$. Average B" scores ($M$ = -0.372, $SD$ = 0.359) were significantly different than 0, indicating an overall bias towards labeling videos as deepfakes even when they are not, $t(203) = -14.810$, $p < .001$, $d = -1.037$, $BF_{10} > 1000$.

There was a significant difference on B" scores as a function of attitudes towards President Biden, $F(6, 197) = 2.734$, $p = 0.014$, $\eta^2 = 0.077$, $BF_{10} = 0.587$ (Figure 5). A posthoc analysis revealed that this was driven by significant differences between those who reported that they liked President Biden "a great deal" and "a little" ($p_{tukey} = 0.010$, $BF_{10, U}$ = 25.664), between those that like President Biden "a great deal" and "dislike a little" ($p_{tukey}$ = 0.016, $BF_{10, U}$ = 41.344), and between those that like "a great deal" those that dislike him "a moderate amount" ($p_{tukey}$ = 0.048, $BF_{10, U}$ = 7.552). No other pairwise comparisons were significant. There was no significant difference between self-identified political ideology on B" scores, $F(6, 197) = 1.463$, $p = 0.193$, $\eta^2 = 0.043$, $BF_{10}$ = 0.053. There was no significant difference between political party affiliation on B" scores, $F(2, 201) = 2.750$, $p = 0.066$, $\eta^2 = 0.027$, $BF_{10}$ = 0.582 (Figure 6).

**Figure 5**

*Bias scores by views towards President Biden before the videos.*

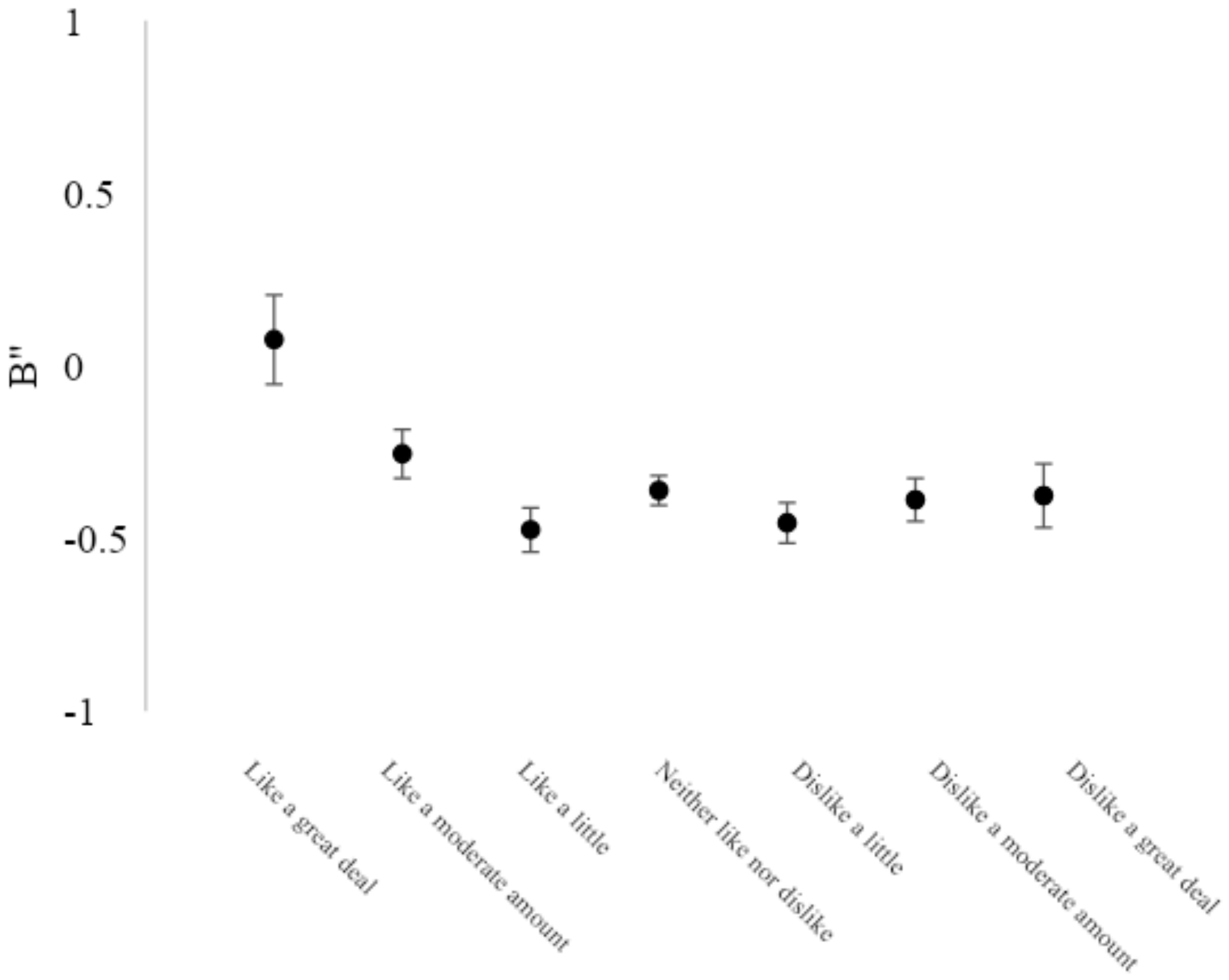

*Note.* Error bars represent ±1 SEM.

**Figure 6**

*Bias scores by political ideology.*

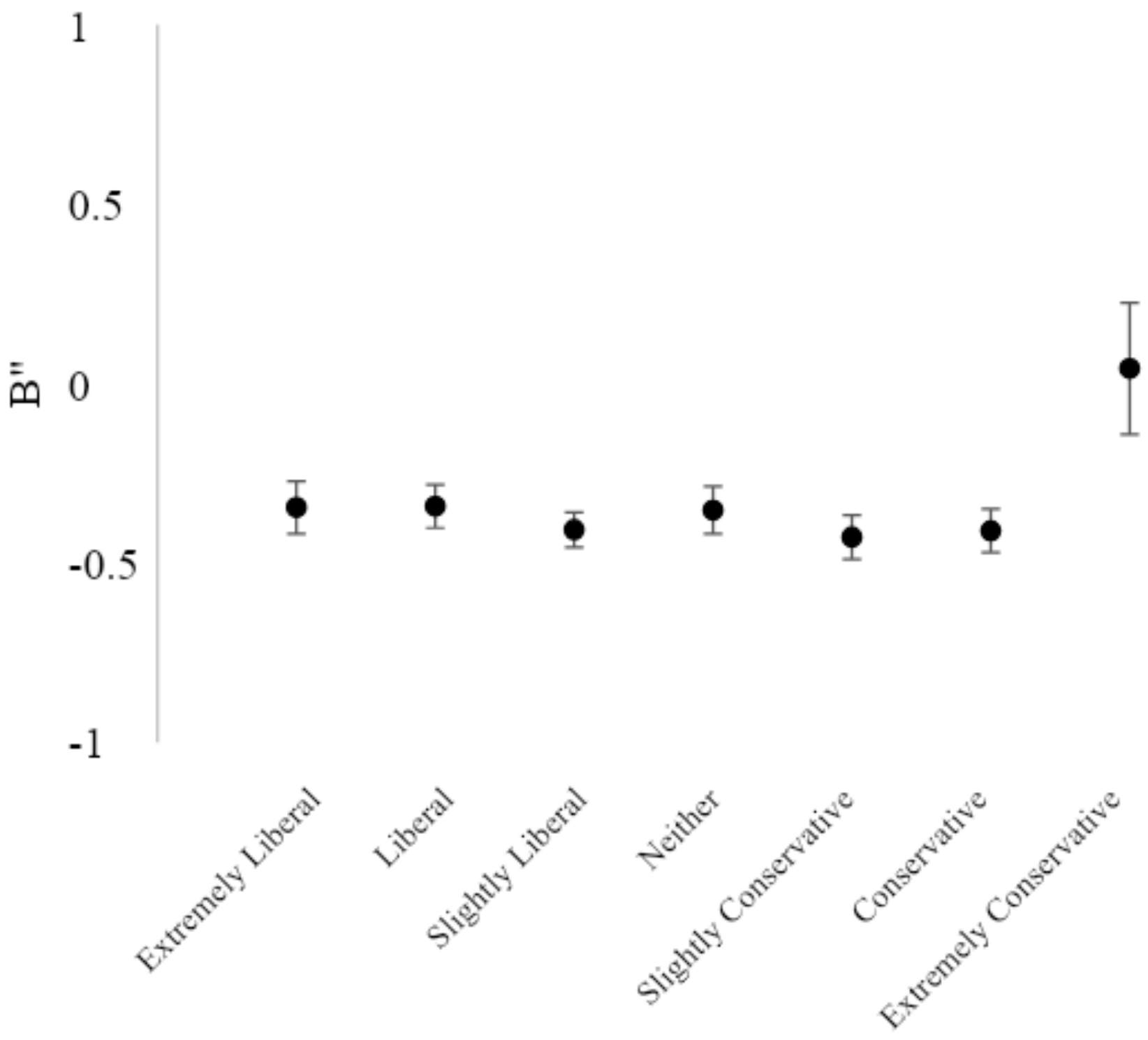

*Note.* Error bars represent ±1 SEM.

**Believability**

Average believability scores (i.e. how believable did participants find the statements) were calculated for each participant. There was a significant difference between political party affiliation on average believability scores, $F(2, 201) = 6.559$, $p = .002$, $\eta^2 = 0.061$, $BF_{10} = 14.531$. A Tukey's post hoc comparison yielded a significant difference between Democrats and Republicans, with Republicans giving higher believability scores ($p_{tukey} = .005$, $BF_{10,\, U} = 9.178$), and between Independents and Republicans, again with Republicans giving higher believability scores ($p_{tukey} = .004$, $BF_{10, U} = 40.561$), however there was no difference between Democrats and Independents ($p_{tukey} = 1.000$, $BF_{10, U} = 0.177$).

**Figure 7**

*Average self-reported believability scores on deepfake video statements by party affiliation*

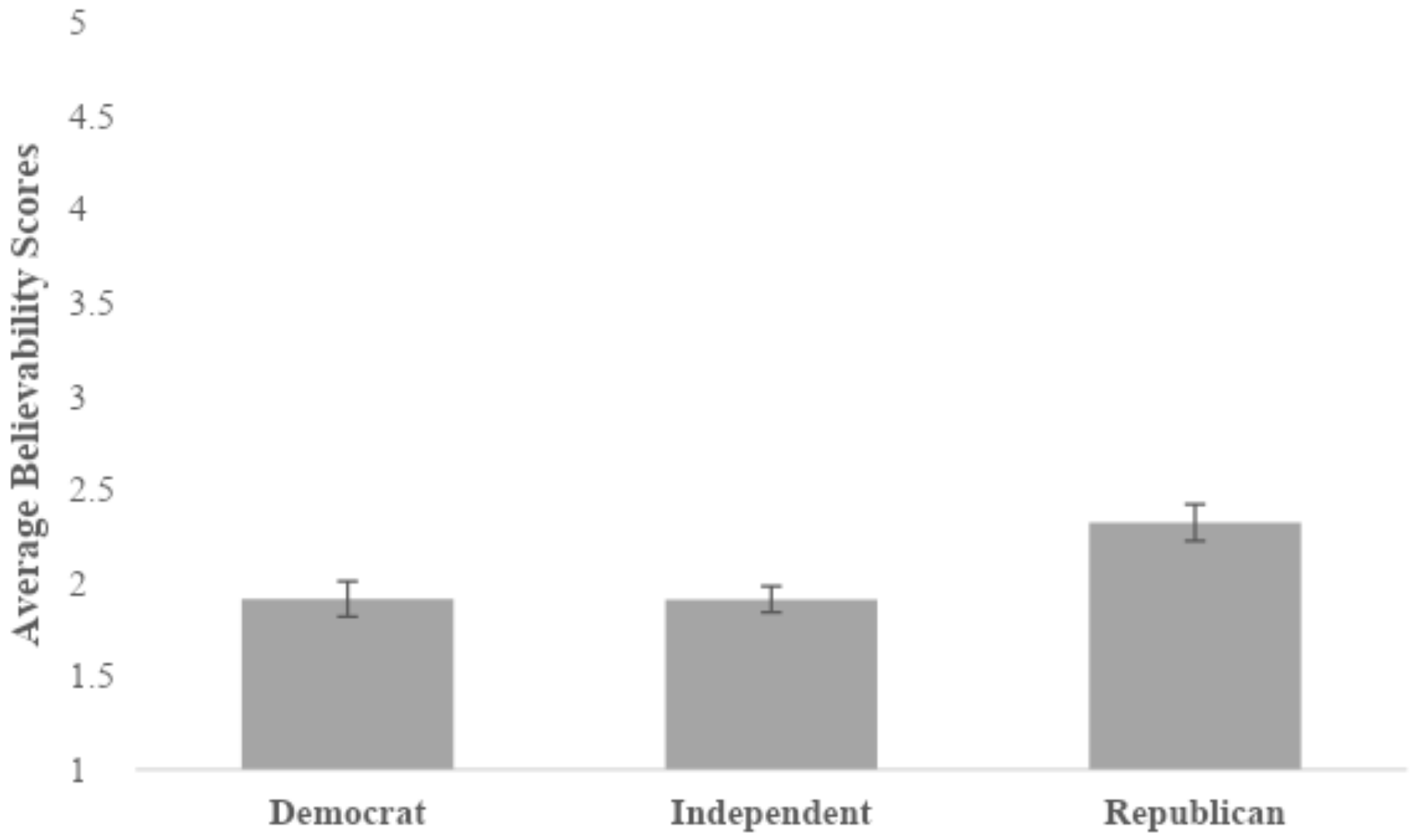

*Note.* Error bars represent ±1 SEM.

There was a significant difference between political ideology on the believability scores, $F(6, 197) = 9.681$, $p < .001$, $\eta^2 = 0.228$, $BF_{10} > 1000$. Tukey's post hoc tests are reported below in Table 6. Comparisons not reported in the table were not significant.

**Table 6**

*Posthoc comparisons for believability ratings*

| | | |
|---|---|---|
| Extremely Conservative | Slightly Conservative | $p_{tukey} < .001$, $BF_{10, U} = 356.251$ |
| | Neither Liberal nor Conservative | $p_{tukey} < .001$, $BF_{10, U} = 128.198$ |
| | Slightly Liberal | $p_{tukey} < .001$, $BF_{10, U} > 1000$ |
| | Liberal | $p_{tukey} < .001$, $BF_{10, U} > 1000$ |
| | Extremely Liberal | $p_{tukey} < .001$, $BF_{10, U} = 452.455$ |
| Conservative | Slightly Conservative | $p_{tukey} = .023$, $BF_{10, U} = 11.959$ |
| | Neither Liberal nor Conservative | $p_{tukey} = .007$, $BF_{10, U} = 15.092$ |
| | Slightly Liberal | $p_{tukey} < .001$, $BF_{10, U} > 1000$ |
| | Liberal | $p_{tukey} < .001$, $BF_{10, U} = 499.012$ |
| | Extremely Liberal | $p_{tukey} < .001$, $BF_{10, U} = 228.856$ |

**Figure 8**

*Average believability scores on deepfake video statements by political ideology.*

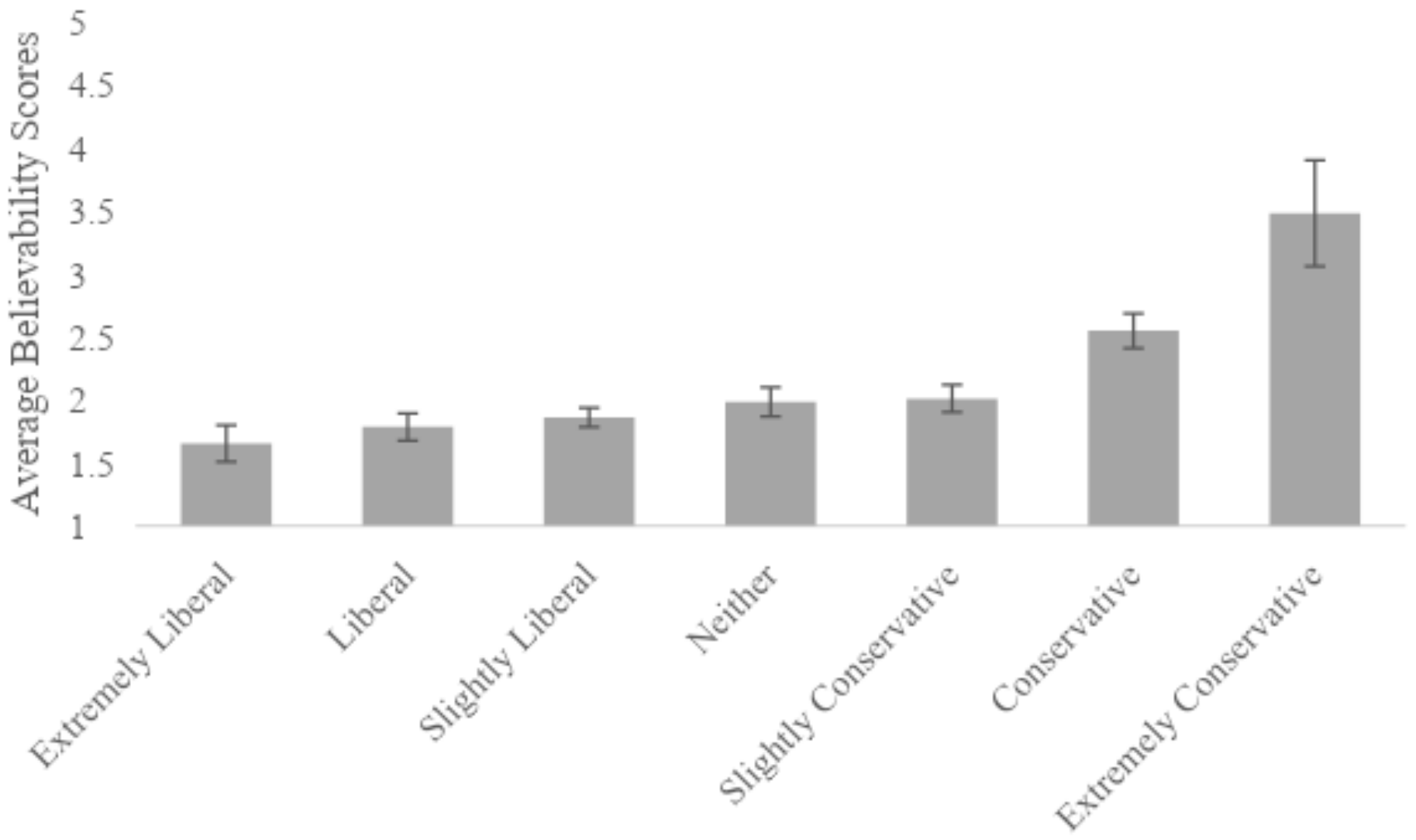

*Note.* Error bars represent ±1 SEM.

Participant views on Biden had a significant effect on believability scores, $F(6, 197) = 5.597$, $p < .001$, $\eta^2 = 0.146$, $BF_{10} = 928.282$. A Tukey's post hoc comparison yielded a significant difference between those who reported disliking Biden "a great deal" and "a little" ($p_{tukey} = .007$, $BF_{10,\,U} = 212.905$), dislike a great deal and neither like nor dislike ($p_{tukey} < .001$, $BF_{10,\,U} > 1000$), dislike a great deal and like a little ($p_{tukey} = .013$, $BF_{10,\,U} = 24.485$), dislike a great deal and like a moderate amount ($p_{tukey} = .019$, $BF_{10,\,U} = 14.541$), and between disliking a moderate amount and neither like nor dislike ($p_{tukey} = .006$, $BF_{10,\,U} = 118.738$). No other comparisons were significant.

**Figure 9**

*Average believability scores of deepfake video statements by views on President Biden before exposure to the videos*

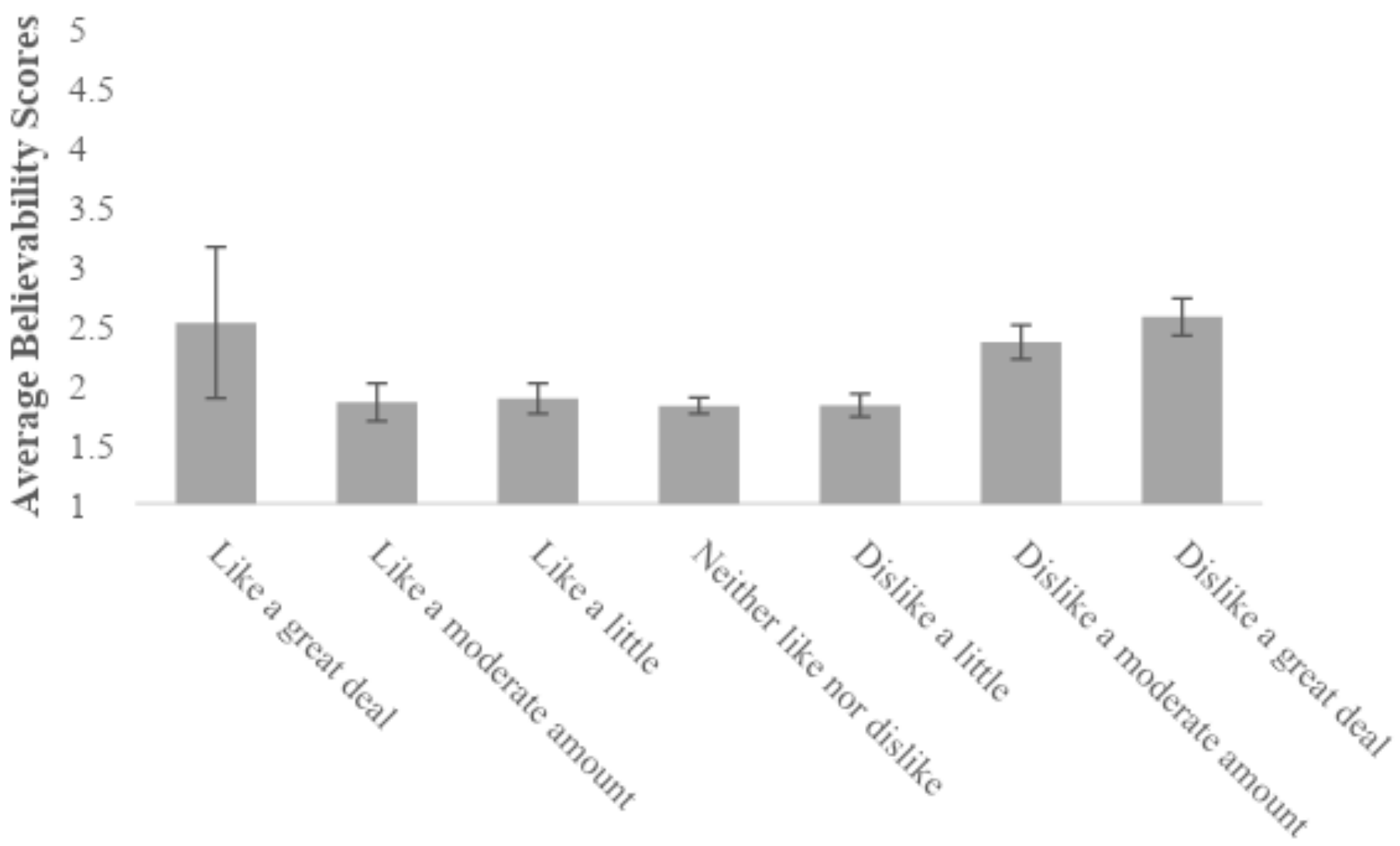

*Note.* Error bars represent ±1 SEM.

There was a significant negative correlation between how believable one thought the deepfakes were on average, and participant sensitivity scores, $r$ (202) = -0.374, $p < .001$, $BF_{10} > 1000$.

**Figure 10**

*Relationship between average believability and sensitivity scores for each subject*

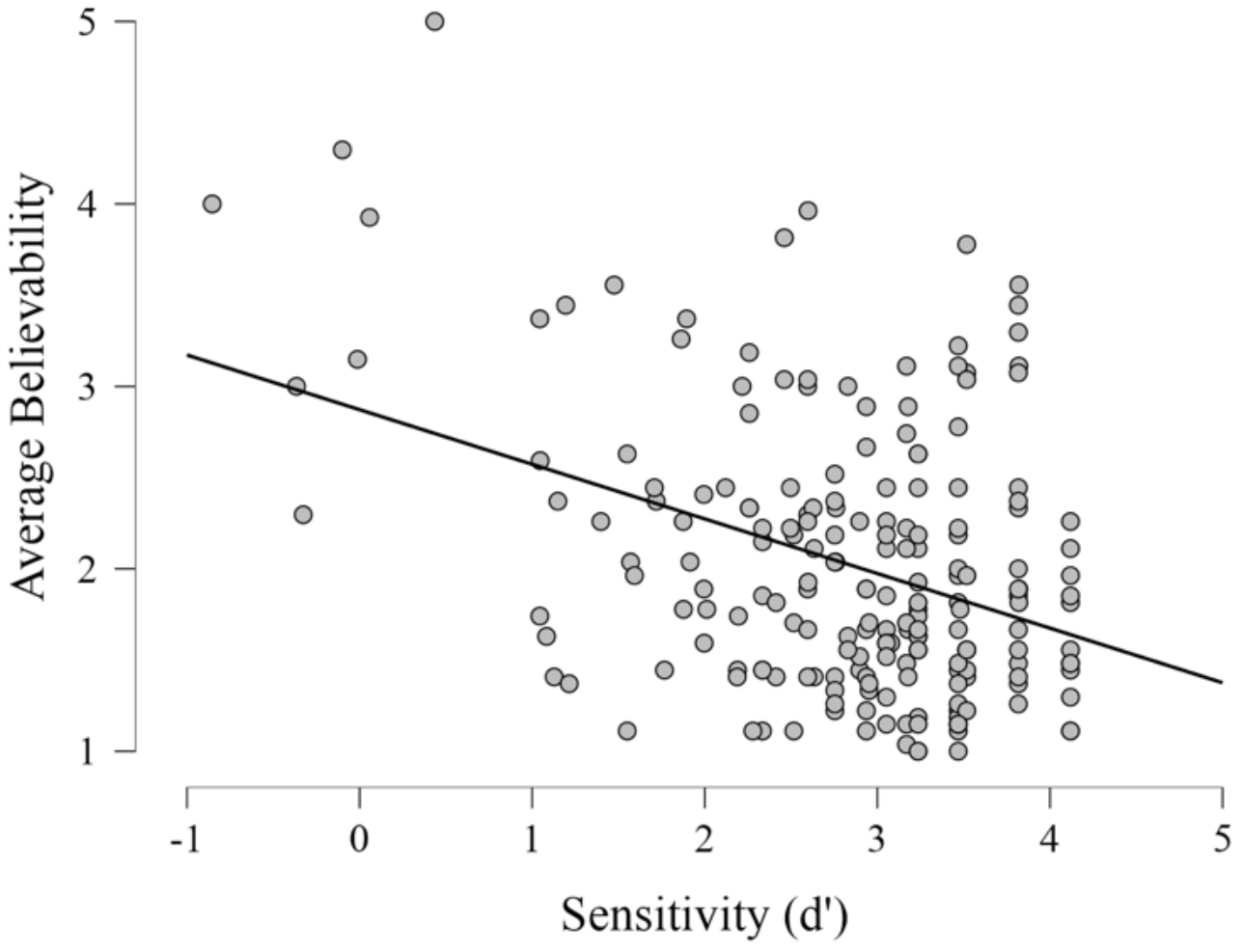

**Table 7**

*Believability Ratings of Deepfake Video Statements by Party Affiliation*

| | Democrat | Independent | Republican |
|---|---|---|---|
| We need to get rid of all fossil fuels, coal, and fracking by the end of my Presidency. | 2.541<br>(1.252) | 2.622<br>(1.224) | 3.054<br>(1.271) |
| Sometimes you have to lie through your teeth to get elected. | 1.635<br>(1.093) | 1.459<br>(0.894) | 1.875<br>(1.176) |
| I don't really like Kamala, I had to pick her to get the Black vote. | 1.459<br>(0.954) | 1.311<br>(0.639) | 1.518<br>(1.009) |
| If the DOJ open an investigation into my son Hunter, I will stop it immediately. | 1.770<br>(1.105) | 1.622<br>(1.030) | 2.375<br>(1.214) |
| I am asking Congress to federalize Planned Parenthood. | 2.703<br>(1.382) | 2.811<br>(1.178) | 3.321<br>(1.223) |
| Cops should automatically be sent to prison if they shoot a black man, even if he is armed. | 1.622<br>(1.143) | 1.608<br>(0.934) | 1.839<br>(1.023) |
| The branches of government should not be equal, the President should have more power than the rest. | 1.419<br>(0.891) | 1.405<br>(0.792) | 1.589<br>(0.949) |
| We need to make DC a state so my party can be in power for 100 years. | 1.419<br>(1.020) | 1.311<br>(0.701) | 1.500<br>(0.915) |
| I am calling for the defunding of every police force in America. | 1.770<br>(1.105) | 1.703<br>(0.887) | 2.232<br>(1.265) |
| There is no place in America for conservative thought. | 1.878 (1.170) | 1.946 (1.071) | 2.393 (1.123) |
| Americans should be wearing masks until at least 2023. | 2.770 (1.245) | 2.811 (1.178) | 3.411 (1.172) |
| By 2030, all cars will have to be electric. | 2.486 (1.295) | 2.622 (1.246) | 2.750 (1.297) |
| We will soon require anyone who owns an AR-15 either give them up or go to jail. | 1.797 (1.193) | 2.041 (1.116) | 2.625 (1.329) |
| If I can make American more socialist than it is today I will have done my job. | 1.838 (1.182) | 1.824 (1.052) | 1.946 (1.086) |
| If the opioid crisis affected minority communities, I would care more about it. | 1.500 (1.037) | 1.486 (0.880) | 1.911 (1.032) |
| Nobody in the United States should own anything other than a musket. | 1.676 (1.148) | 1.622 (0.932) | 2.214 (1.303) |
| We need to do away with all religious, private schools. | 1.676 (1.022) | 1.608 (0.841) | 2.071 (1.126) |
| Open borders mean Democrats never lose another election and that is good for America. | 1.568 (0.980) | 1.432 (0.812) | 1.946 (1.151) |
| The pandemic has given us the perfect opportunity to push our agenda. | 1.703 (1.144) | 1.473 (0.864) | 2.232 (1.191) |
| I think schools need to be closed for at least another year. | 2.662 (1.174) | 2.959 (1.232) | 3.196 (1.242) |
| I plan on stepping down in a year or two and let Kamala take over as President. | 1.851 (1.155) | 1.649 (0.971) | 1.911 (1.032) |
| We need to raise the tax rate on the top and middle class to at least 70%. | 1.851 (1.131) | 1.946 (1.169) | 2.732 (1.328) |
| Frankly, I think Texas should succeed [SIC]. | 1.581 (1.102) | 1.608 (0.873) | 1.946 (1.135) |
| Welfare should be for those who can't but also don't want to work. | 1.703 (1.069) | 1.770 (0.944) | 2.179 (1.336) |
| If another new variant of COVID-19 comes to America I will lock us down immediately. | 2.905 (1.218) | 2.851 (1.106) | 3.161 (1.262) |
| I would be in favor of changing the name of the White House to something more politically correct. | 1.946 (1.215) | 2.162 (1.250) | 2.661 (1.311) |
| Rap music is the reason so many young people are getting into crime these days. | 1.851 (1.201) | 1.865 (1.089) | 2.018 (1.183) |

## Confidence & Scrutiny

Participants believe themselves to be quite good at detecting deepfakes in general, with an average just over the 76th percentile ($M$ = 76.265, $SD$ = 18.164). Participants also believed themselves to be very good in the present study, with an average nearly in the 73rd percentile ($M$ = 72.936, $SD$ = 19.639). Participants also believed themselves to be significantly better on this task in terms of correct recall ($M$ =

39.662, *SD* = 9.235) than their peers in the study (*M* = 35.520, *SD* = 9.097), *t*(203) = 8.466, *p* < .001, *d* = 0.593, $BF_{10}$ > 1000. While participants, on average, overestimated their ability relative to their peers, they actually significantly underestimated their actual recall performance (*M* = 45.539, *SD* = 5.572), *t*(204) = 9.993, *p* < .001, *d* = 0.700, $BF_{10}$ > 1000. There was a significant correlation between the number of videos that participants correctly recalled, and the number that they guessed they would correctly recall, *r*(202) = 0.445, *p* < .001, $BF_{10}$ > 1000.

After viewing all of the videos, participants were asked to complete a scrutiny scale to serve as an attention check for the videos. The use of scrutiny as an attention/manipulation check is consistent with prior research examining deepfakes (Dobber, 2021). We used the three scrutiny items utilized by Dobber, based on a four-item scale originally developed by Wheeler and colleagues (2005). It should be noted that the scrutiny scale we used in our study was a four-item scale with a fifth response option of "I don't know" while Wheeler and colleagues' original scrutiny measure included a 7-item scale. In the current study, we deployed the scrutiny scale as an attention check for the videos participants viewed. Because prior research in the ELM literature suggests that outgroup members will scrutinize information more closely than ingroup members, we conducted the following post hoc analyses. We examined the relationship between liking President Biden and video scrutiny scores. There was no significant difference between how much someone reported liking President Biden at the start of the experiment, on the self-reported level of scrutiny given to the videos, *F*(6, 197) = 1.750, *p* = .111, $BF_{10}$ = 0.184. Similarly, there was no significant difference between political ideology on the self-reported level of scrutiny given to the videos, $F(6, 197) = 1.108$, $p = .359$, $BF_{10} = 0.087$.

## Discussion

In the current study, we presented participants with both deepfake and actual videos with accurate labels. Participants were then given a series of questionnaires, after which they were presented with each statement made in the videos and asked to recall whether the statement came from a deepfake or an actual video. Participant sensitivity was relatively high for this task. Overall, participants recalled 93.8% of deepfake videos and 84.2% of actual videos. Participant performed much better in recalling and identifying labeled deepfake videos than they perform in deepfake identification studies that do not use content labels (Barari et. al., 2021; Rosseler et. al.; Appel & Prietzel, 2022; Hughes et. al., 2021), where participants often performed at or worse than levels of chance. Our results indicate that labeling false information can counteract misinformation’s negative effects, similar to Lee and Shin (2022), who found that labeling fake news videos reduced participant’s engagement intentions and increased cognitive guard. Labeling deepfake content on platforms such as social media sites can help people remember which information is false, offering a potentially effective approach to combating disinformation spread through deepfake videos.

While participants recalled deepfake and actual videos at significantly better levels than chance, participants demonstrated a bias towards labeling videos as “deepfakes” even when they were actual. When participants made errors, they were significantly more likely to commit a “false alarm” error by saying that an authentic video was a deepfake than to commit a “miss” by identifying a deepfake as an authentic video. This finding is consistent with other studies in which participants demonstrated a propensity to over-identify videos as deepfakes (Ternovski, Kalla, & Aronow, 2022; Vaccari & Chadwick, 2020). According to error management theory (Ng, 2022), participants are biased towards assuming that genuine videos contain untrue, dangerous content because this error is perceived as less costly.

Although this bias helps explain why people performed better at identifying deepfakes, another reason for this difference in performance could come from the fact that some of the deepfake video statements were blatantly absurd (i.e. “I would be in favor of changing the name of the White House to something more politically correct”). This explanation is further corroborated by evidence suggesting that overall, participants who found the deepfake videos more believable were more likely to label them as actual. As the participants’ reported believability of the statements increased, their sensitivity decreased. This finding aligns with an assumption that those who find deepfake statements to be more believable would more often struggle to distinguish them from actual statements.

However, while the overall trend showed that increased believability reduced sensitivity, individual differences complicate this analysis. When we clustered the participants by political party affiliation (Democrat, Independent, Republican), a pattern emerged, with Republicans rating deepfake statements as significantly more believable. Despite finding the videos more believable as a whole, participants who affiliated with the Republican party were not significantly worse than other participants at recalling whether individual videos were authentic or deepfakes. Even more than political identification, attitudes towards Biden were the biggest predictor in participant ability to correctly recall whether videos were actual or deepfake. In general, participants who had higher favorability ratings for Biden performed worse at recall tasks by more often identifying deepfake videos as actual videos. In contrast, participants with lower perceptions of Biden tended to more accurately recall whether videos were actual or deepfake. Analyzed through the ELM framework, it is not surprising that we found individuals who identify as Republican and those who had lower perceptions of Biden performed better on recall tasks than other participants. ELM suggests that both Republicans and those with lower perceptions of Biden would be more likely to view the message source of our deepfake videos (i.e., Biden) as less trustworthy and this skepticism would increase message elaboration (Petty et al., 2009). Increased elaboration could explain increased recall accuracy among these two groups. This finding aligns with MacDonald, Milfont, and Gavin (2016), who found that messages that prompted more elaboration increased participant recall.

That attitudes towards President Biden was a negative predictor of recall performance in deepfakes is somewhat surprising, especially on the deepfake video recall, given that the deepfakes were often somewhat extreme positions we did not believe were representative of statements President Biden would make, and as such it was expected that those who like President Biden more, may be less willing to believe that he said the extreme statements. Understanding these individual differences may help combat negative influences of deepfakes by helping researchers and media outlets identify individuals most susceptible to deepfake misinformation.

Other individual characteristics affected participants' recall performance, though these characteristics differed for deepfake and actual videos. For the deepfake videos, the number of hours someone self-reported spending reading and watching the news, the believability of the deepfake statements, and one's views on President Biden were all negative predictors of recall performance; as each of those measures increased, recall performance decreased. However, the number of different sources that one engages with was a positive predictor of recall performance (i.e. a greater variety of news sources was predictive of better recall performance for deepfake videos). Similarly, when we looked at factors which predict recall performance on the actual video, we found that the number of hours someone self-reported spending reading and watching the news, and the believability of the deepfake statements were negative predictors of recall performance. It is unknown whether this is causal, and if it is, what direction the causality takes.

Another concern surrounding disinformation is that deepfakes could potentially swing elections by changing beliefs about the candidate. In this present study we found no evidence for a change in attitudes towards President Biden as a function of being exposed to labeled deepfakes. This finding contrasts with previous research that correlated deepfake exposure with lower attitudes about the figures depicted in the videos (Dobber et. al., 2021), and conflicts somewhat with findings that participants will more positively rank political figures after hearing deepfake claims that align with participants' existing views (Hameleers, 2022). While this finding does not necessarily speak to the behavior of the electorate, it does at least suggest that mere exposure to labeled deepfakes may not be sufficient to change perceptions of an individual.

This study relied on participants from a Western, educated, industrialized, rich and democratic (WEIRD) society, which means that results are not generalizable to all populations. However, our participant group best reflects individuals who would naturally see and respond to deepfake videos. The study emphasized individuals in the United States with access to networked technology (computers, smartphones), but individuals with those devices will most frequently encounter deepfakes. Additionally, individuals in democratic societies may exhibit more interest and response to deepfake videos about elected politicians.

## Limitations

While this study used videos and deepfakes of President Biden, we have no reason to believe that our results are unique to President Biden. However, it is important to note that during the course of data collection, President Biden's overall approval numbers have hovered in the mid- to high-30s (REF). It is therefore plausible that results may differ were the data collected earlier on in President Biden's tenure when his approval ratings were in the mid-50s.  The current study created deepfake videos using some of the earliest software available to users to create such content. We degraded real videos so all videos appeared to the viewer to be equivalent in quality but as deepfake technology improves it would be useful to replicate this study to determine if higher quality deepfake videos might impact the results. Another limitation stems from our use of the working memory capacity measure. It is unclear why so many participants had a zero for this score; we are inclined to believe there was either an idiosyncrasy in the way these were coded to be presented to the users, or perhaps the instructions were insufficiently clear. As such, caution should be exerted when interpreting the results as they relate to working memory.

## Conclusion

In the present study, participants could remember with high accuracy when statements came from videos labeled as either “actual” or “deepfake.” However, participants were significantly better at recalling videos labeled as deepfake correctly than they were at recalling videos labeled as actual videos. This finding suggests that labeling fake videos as deepfakes may be effective in preventing that information from being encoded and recalled as true information. Participants with lower approval ratings of Biden showed more sensitivity to remembering that content came from a deepfake video, suggesting that individuals who favor a speaker in a deepfake video may be more vulnerable to its influence.

## Appendix A

Deepfake and Actual Statements Made by Joe Biden in Experimental Videos

| Deepfake | Actual |
|---|---|
| We need to get rid of all fossil fuels, coal, and fracking by the end of my Presidency. | I announced the United States made a new commitment under the Paris Agreement to cut our emissions by 50 – 52% by the end of this decade. |
| Sometimes you have to lie through your teeth to get elected. | And each of us has a duty and a responsibility, as citizens, as Americans, and especially as leaders, leaders who have pledged to honor our constitution, and protect our nation. |
| I don't really like Kamala, I had to pick her to get the Black vote. | She’s doing a great job. She’s a great partner. |
| If the DOJ open an investigation into my son Hunter, I will stop it immediately. | |
| I am asking Congress to federalize Planned Parenthood. | I would codify Roe v. Wade, and Casey. |
| Cops should automatically be sent to prison if they shoot a black man, even if he is armed. | We need criminal justice reform, but that isn’t nearly enough. |
| The branches of government should not be equal, the President should have more power than the rest. | We have the tools to help people, so let’s use the tools, all of them, use them now. |
| We need to make DC a state so my party can be in power for 100 years. | I’m a proud Democrat, and I will be proud to carry the banner of our party into the general election. |
| I am calling for the defunding of every police force in America. | The vast majority of men and women wearing a uniform and a badge serve our communities and they serve them honorably. |
| There is no place in America for conservative thought. | We must end this uncivil war, that pits red against blue. |
| Americans should be wearing masks until at least 2023. | We need everyone to get vaccinated. We need everyone to keep washing their hands. |
| By 2030, all cars will have to be electric. | We are going to boost our manufacturing capacity. That's why the American Jobs Plan invests in new and retooled union facilities, grants to kickstart new battery and parts production |
| We will soon require anyone who owns an AR-15 either give them up or go to jail. | We need a ban on assault weapons and high capacity magazines. |
| If I can make American more socialist than it is today I will have done my job. | The difference between the President and I think healthcare is not a privilege, it's a right. |
| If the opioid crisis affected minority communities, I would care more about it. | Addiction touches families in every community, in red states and blue states, across all races and creeds |
| Nobody in the United States should own anything other than a musket. | In the very beginning you couldn't own any weapon you wanted to own. |
| We need to do away with all religious, private schools. | That's why my American families plan guarantees four additional years of public education for every person in America. |
| Open borders mean Democrats never lose another election and that is good for America. | Let's end our exhausting war over immigration |
| The pandemic has given us the perfect opportunity to push our agenda. | Shut down the virus, not the country. |
| I think schools need to be closed for at least another year. | Seeing the smiles on the kids' faces for those who are able to go back to school because the teachers and school bus drivers and cafeteria workers have been vaccinated. |
| I plan on stepping down in a year or two and let Kamala take over as President. | Yes, my plan is to run for re-election, that's my expectation. |
| We need to raise the tax rate on the top and middle class to at least 70%. | We take the top tax bracket, for the wealthiest 1% of Americans, those making over $400,000 or more, back up to where it was when George W. Bush was President |
| Frankly, I think Texas should succeed [SIC]. | |
| Welfare should be for those who can't but also don't want to work. | If you're receiving unemployment benefits, and you're offered a suitable job, you can't refuse that job, and just keep getting the unemployment benefits. |
| If another new variant of COVID-19 comes to America I will lock us down immediately. | Because if we don't stay vigilant, and the conditions change, then we may have to reinstate restrictions to get back on track. |
| I would be in favor of changing the name of the White House to something more politically correct. | |
| Rap music is the reason so many young people are getting into crime these days. | There's no place for violence |
| | White supremacy is terrorism. We're not going to ignore that either. |